\documentstyle[twocolumn,seceq,epsbox]{jpsj}

\def\gsim{\,$\raise0.3ex\hbox{$>$}\llap{\lower0.8ex\hbox{$\sim$}}$\,}
\def\lsim{\,$\raise0.3ex\hbox{$<$}\llap{\lower0.8ex\hbox{$\sim$}}$\,}
\def\lrarrow{\,$\raise0.4ex\hbox{$\rightarrow$}\llap{\lower0.4ex\hbox{$\left
arrow$}}$\,}

\def\vecS{{\vec S}}
\def\vecT{{\vec T}}
\def\bE{{\bar E}}
\def\a{\alpha}
\def\b{\beta}

\def\runtitle{Ground State Properties of an $S\!=\!1/2$ Distorted Diamond
Chain}

\def\runauthor{T.~{\sc Tonegawa}, K.~{\sc Okamoto}, T.~{\sc Hikihara},
Y.~{\sc Takahashi} and M.~{\sc Kaburagi}}

\title
{
Ground State Properties of an $S\!=\!1/2$ Distorted Diamond Chain
}

\author
{
Takashi {\sc Tonegawa}\footnote{tonegawa@kobe-u.ac.jp},
Kiyomi {\sc Okamoto}$^{1,}$\footnote{kokamoto@stat.phys.titech.ac.jp},
Toshiya {\sc Hikihara}$^{2}$, Yutaka {\sc Takahashi}$^{3}$, and
Makoto {\sc Kaburagi}$^{4}$
}

\inst
{
Department of Physics, Faculty of Science, Kobe University, Rokkodai,
Kobe 657-8501\\
$^{1}$Department of Physics, Tokyo Institute of Technology, Oh-Okayama,
Meguro-ku, Tokyo 152-8551\\
$^{2}$ Division of Information and Media Science, Graduate School of Science
and Technology, Kobe University, Rokkodai, Kobe 657-8501\\
$^{3}$ Division of Physics, Graduate School of Science and Technology, Kobe
University, Rokkodai, Kobe 657-8501\\
$^{4}$Department of Informatics, Faculty of Cross-Cultural Studies,
Kobe University, Tsurukabuto, Kobe 657-8501
}

\recdate
{December 3, 1999
}

\abst
{We investigate the ground state properties of an \hbox{$S\!=\!1/2$} distorted
diamond chain described by the Hamiltonian
${\cal H}\!=\!J_1\sum_\ell\bigl\{\bigl(\vecS_{3\ell-1}\cdot\vecS_{3\ell}
\!+\!\vecS_{3\ell}\cdot\vecS_{3\ell+1}\bigr)\!+\!J_2\vecS_{3\ell-2}\cdot\vecS_{3\ell-1}
\!+\!J_3\bigl(\vecS_{3\ell-2}\cdot\vecS_{3\ell}\!+\!\vecS_{3\ell}\cdot\vecS_{3\ell+2}\bigr)\!-\!H S_{\ell}^z\bigr\}$ ($J_1,~J_2,~J_3\!\geq\!0$),
which models well a trimerized $S\!=\!1/2$ spin chain system
Cu$_3$Cl$_6$(H$_2$O)$_2$$\cdot$2H$_8$C$_4$SO$_2$.  Using an exact
diagonalization method by means of the Lancz{\"o}s technique, we determine the
ground state phase diagram in the \hbox{$H\!=\!0$} case, composed of the
dimerized, spin fluid, and ferrimagnetic phases.  Performing a degenerate
perturbation calculation, we analyze the phase boundary line between the
latter two phases in the \hbox{$J_2,~J_3\!\ll\!J_1$} limit, the result of
which is in good agreement with the numerical result.  We calculate, by the
use of the density matrix renormalization group method, the ground state
magnetization curve for the case~(a) where $J_1\!=\!1.0$, $J_2\!=\!0.8$, and
$J_3\!=\!0.5$, and the case~(b) where $J_1\!=\!1.0$, $J_2\!=\!0.8$, and
$J_3\!=\!0.3$.   We find that in the case~(b) the $2/3$-plateau appears in
addition to the $1/3$-plateau which also appears in the case~(a).  The
translational symmetry of the Hamiltonian ${\cal H}$ is spontaneously broken
in the $2/3$-plateau state as well as in the dimerized state.
}

\kword
{$S\!=\!1/2$ distorted diamond chain,
Cu$_3$Cl$_6$(H$_2$O)$_2$$\cdot$2H$_8$C$_4$SO$_2$, ground state phase diagram,
dimerized phase, spin fluid phase, ferrimagnetic phase, ground state
magnetization curve, magnetization plateau, spontaneous symmetry breaking,
exact diagonalization method, density matrix renormalization group method,
method of level spectroscopy
}

\begin{document}
\sloppy
\maketitle

\section{Introduction}

In the past years a great deal of work has been devoted to the study of
quantum spin systems with competing interactions.  This has stemmed largely
from their being systems which include two phenomena of great interest,
frustration and quantum fluctuation.  Recently, magnetic properties of a
trimerized $S\!=\!1/2$ spin chain system
Cu$_3$Cl$_6$(H$_2$O)$_2$$\cdot$2H$_8$C$_4$SO$_2$ have been experimentally
studied by Ishii, Tanaka, Hori, Uekusa, Ohashi, Tatani, Narumi, and
Kindo.~\cite{rf:1} (See below for the discussion on the competition of the
exchange interactions in this system.)  They have measured the temperature
dependence of the magnetic susceptibility and also the magnetization curve at
sufficiently low temperatures.  Their results clearly demonstrate that the
ground state of this system is nonmagnetic and that there exists the energy
gap with the magnitude of about $3.9\,{\rm T}$ or $5.2\,{\rm K}$ between the
ground state and a first exited magnetic state.  Furthermore, the result for
the magnetization measurement shows that the so-called 1/3-plateau in the
magnetization curve may start from about $58\,{\rm T}$.

As has been discussed by Ishii {\it et} {\it al.},~\cite{rf:1} the magnetic
properties of Cu$_3$Cl$_6$(H$_2$O)$_2$$\cdot$2H$_8$C$_4$SO$_2$ can be
well described by the following Hamiltonian ${\cal H}$, which is a sum of the
exchange interaction term ${\cal H_{\rm ex}}$ and the Zeeman term
${\cal H_{\rm Z}}$:

\begin{subeqnarray}
 \hspace{-0.50truecm}
 {\cal H} =&&\hspace{-0.80truecm} {\cal H_{\rm ex}} + {\cal H_{\rm Z}}\;, \\
    \hspace{-0.50truecm}
    &&\hspace{-0.80truecm}
      {\cal H_{\rm ex}}
          = J_1 \sum_{\ell=1}^{N/3}
                \bigl(\vecS_{3\ell-1} \cdot \vecS_{3\ell}
                + \vecS_{3\ell}   \cdot \vecS_{3\ell+1} \bigr)  \nonumber  \\
       \hspace{-0.50truecm}
       && + J_2 \sum_{\ell=1}^{N/3}
                \vecS_{3\ell-2} \cdot \vecS_{3\ell-1}           \nonumber  \\
       \hspace{-0.50truecm}
       && + J_3 \sum_{\ell=1}^{N/3}
                \bigl(\vecS_{3\ell-2} \cdot \vecS_{3\ell}
                + \vecS_{3\ell}   \cdot \vecS_{3\ell+2} \bigr)\;,          \\
    \hspace{-0.50truecm}
    &&\hspace{-0.80truecm}
      {\cal H_{\rm Z}} = -H \sum_{\ell=1}^{N} S_{\ell}^z\;,
\end{subeqnarray}
\noindent
where $\vecS_{\ell}$ is the spin operator with the magnitude $S\!=\!1/2$
for the Cu$^{2+}$ ion located at the $\ell$th site, and periodic boundary
conditions $\bigl(\vecS_{N+1}\!\equiv\!\vecS_{1},
\vecS_{N+2}\!\equiv\!\vecS_{2}\bigr)$ are imposed.  In the following
discussions, the total number $N$ of spins is assumed to be a multiple of
six.  From the crystal structure and the lattice parameters of
Cu$_3$Cl$_6$(H$_2$O)$_2$$\cdot$2H$_8$C$_4$SO$_2$, Ishii {\it et}
{\it al.},~\cite{rf:1} have inferred that three exchange constants, $J_1$,
$J_2$, and $J_3$ are all antiferromagnetic (positive).  Thus, the three
interactions compete with each other.  They have also inferred that $J_1$ is
larger than $J_2$ and $J_3$.  This means that the spins $\vecS_{3\ell-1}$,
$\vecS_{3\ell}$, and $\vecS_{3\ell+1}$ form a trimer; $J_1$ is the
intra-trimer exchange constant, and $J_2$ and $J_3$ are the inter-trimer
ones.  Schematical representations of the Hamiltonian ${\cal H_{\rm ex}}$ are
given in Figs.$\,$1(a) and 1(b).   Referring to Fig.$\,$1(a), we may call
this model the ^^ distorted diamond chain model'.

\begin{figure}
  \begin{center}
    \psbox[width=7.8cm]{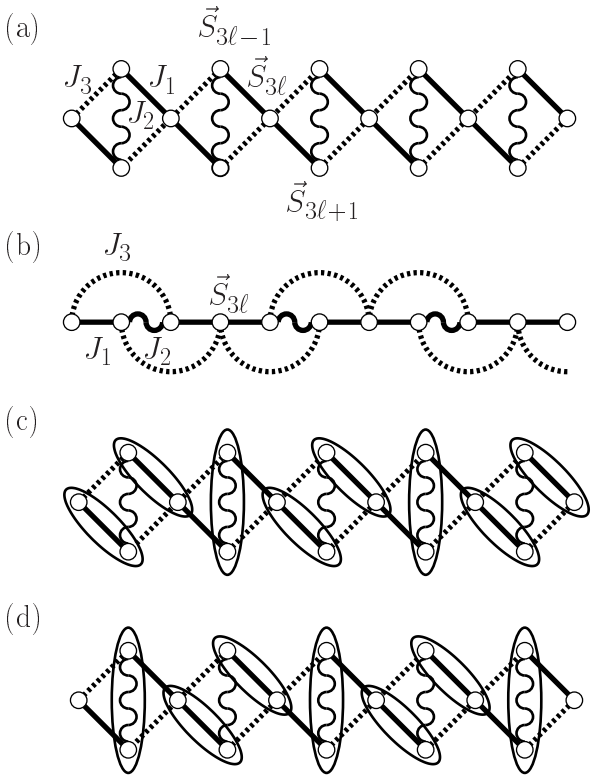}
  \end{center}
\caption{(a) and (b) Two ways of representing schematically the Hamiltonian
${\cal H_{\rm ex}}$.  (c) and (d) The two-fold degenerate dimerized (D)
states.  In (a)$\,$-$\,$(d), the open circles stand for the $S\!=\!1/2$ spins,
and the solid, wavy, and dotted lines correspond to the $J_1$, $J_2$, and
$J_3$ interactions, respectively.  Two spins surrounded by an ellipse in (c)
and (d) form a singlet dimer}
\end{figure}

In the present paper, we explore the ground-state properties of the distorted
diamond chain, performing mainly numerical calculations based on the
Lancz{\"o}s method as well as the density matrix renormalization group (DMRG)
method.~\cite{rf:2,rf:3}  We focus our attention upon the ground state phase
diagram in the case of zero external magnetic field ($H\!=\!0$) and the ground
state magnetization curve.  We treat the case where \hbox{$J_1\!\geq\!J_3$}
and \hbox{$J_2\!\geq\!0$}, noting that the \hbox{$J_1\!<\!J_3$} case is
equivalent to the \hbox{$J_1\!>\!J_3$} case by interchanging the role of $J_1$
and $J_3$.

Let us now discuss three special cases where $H\!=\!0$ is assumed.  First,
when $J_2\!=\!0$, we can readily show by the use of the Lieb-Mattis
theorem~\cite{rf:4} that the ground state is a ferrimagnetic (FRI) state
where the magnitude $S_{\rm tot}$ of the total spin
$\vecS_{\rm tot}\!=\!\sum_{\ell=1}^{N}\vecS_{\ell}$
$\bigl(\vecS_{\rm tot}^2\!=\!S_{\rm tot}(S_{\rm tot}\!+\!1)\bigr)$ is given
by $S_{\rm tot}\!=\!N/6$.  We note that this value is one third of that of
the maximum (ferromagnetic) value, $N/2$.  Second, when $J_3\!=\!0$, the
system is reduced to the $J_1$-$J_1$-$J_2$ trimerized $S\!=\!1/2$
antiferromagnetic chain; in particular, if $J_1\!=\!J_2$, the $S\!=\!1/2$ 
uniform antiferromagnetic chain is obtained.  Thus, the ground state is
the spin fluid (SF) state with \hbox{$S_{\rm tot}\!=\!0$}, which has no
excitation energy gap.~\cite{rf:5}

Third, the case where $J_1\!=\!J_3$ has already been investigated by Takano,
Kubo, and Sakamoto.~\cite{rf:6}  They have shown that the ground state is
the FRI state with $S_{\rm tot}\!=\!N/6$, the tetramer-dimer (TD) state with
$S_{\rm tot}\!=\!0$, or the dimer-monomer (DM) state with $S_{\rm tot}\!=\!0$,
depending on whether $J_2/J_1\!<\!0.909$, $0.909\!<\!J_2/J_1\!<\!2$, or
$2\!<\!J_2/J_1$.  In the TD state, quadruplets $\vecS_{3\ell-3}$,
$\vecS_{3\ell-2}$, $\vecS_{3\ell-1}$, and $\vecS_{3\ell}$ (or $\vecS_{3\ell}$,
$\vecS_{3\ell+1}$, $\vecS_{3\ell+2}$, and $\vecS_{3\ell+3}$) of spins form
singlet tetramers, and pairs $\vecS_{3\ell+1}$ and $\vecS_{3\ell+2}$
($\vecS_{3\ell+4}$ and $\vecS_{3\ell+5}$) of spins, which we call $J_2$-pairs,
singlet dimers.  Thus, the TD state is two-fold degenerate.  In the DM state,
on the other hand, the $J_2$-pairs form singlet dimers, and the remaining
monomer spins $\vecS_{3\ell}$ are free, which leads to $2^{N/3}$-fold
degeneracy.  These are schematically illustrated in Fig.$\,$1 in
ref.$\,$6.  The explicit expressions of the eigenfunctions and some physical
quantities for both the TD and the DM state are given in Appendix.  As has
been shown by Takano ${\it et}$ ${\it al.}$,~\cite{rf:6} the commutation
relation
$\big[{\cal H_{\rm ex}},\;\bigl(\vecS_{3\ell+1}\!+\!\vecS_{3\ell+2}\bigr)^2\bigr]\!=\!0$ holds in the $J_1\!=\!J_3$ case.  Therefore, each $J_2$-pair forms a
singlet or a triplet.  It is interesting to note that in the DM and FRI
states, all the $J_2$-pairs form the singlets and the triplets, respectively,
while in the TD state the singlet and triplet $J_2$-pairs are arranged
alternatively.

In a recent paper~\cite{rf:7}, Okamoto, Tonegawa, Takahashi, and Kaburagi
have studied the ground state of the distorted diamond chain in the case of
$H\!=\!0$, employing analytical methods based mainly on the bosonization
technique and doing physical considerations.  According to them, a tetramer
in Takano ${\it et}$ ${\it al.}$'s TD state is very peculiar to the
\hbox{$J_1\!=\!J_3$} case, and when \hbox{$J_1\!>\!J_3$}, the tetramer
consisting of the spins, $\vecS_{3\ell-3}$, $\vecS_{3\ell-2}$,
$\vecS_{3\ell-1}$, and $\vecS_{3\ell}$ is decomposed into two dimers, one of
which consists of $\vecS_{3\ell-3}$ and $\vecS_{3\ell-2}$, and the other of
$\vecS_{3\ell-2}$ and $\vecS_{3\ell-1}$.  Thus, the TD state is a special
case of the two-fold degenerate dimerized (D) state with
\hbox{$S_{\rm tot}\!=\!0$}, shown in Figs.$\,$1(c) and 1(d).  The DM state
is also peculiar to the \hbox{$J_1\!=\!J_3$} case.  When \hbox{$J_1\!>\!J_3$},
the monomer is no longer free and interacts with neighboring monomers through
the dimer between them.  Thus, the DM state is a special case of the SF state
with \hbox{$S_{\rm tot}\!=\!0$}.

Summarizing these, Okamoto ${\it et}$ ${\it al.}$~\cite{rf:7} have concluded
that the ground state phase diagram of the distorted diamond chain is composed
of the D and SF phases plus the FRI phase which is stable at least when $J_2$
is sufficiently smaller than $J_1$ and $J_3$.  They have also shown that the
transition between the D and SF phases is of the
Berezinskii-Kostelitz-Thouless (BKT) type, as in the case of the
\hbox{$S\!=\!1/2$} antiferromagnetic chain with uniform nearest and
next-nearest neighbor interactions,~\cite{rf:8} and have presented the ground
state phase diagram determined numerically.  In the region where the D state
is the ground state, there exists a finite gap between the two-fold degenerate
ground state and a first exited state.  As can be seen from Figs.$\,$1(c) and
1(d), the period of the translational symmetry of the D state is six, which is
twice as large as that of the Hamiltonian ${\cal H}_{\rm ex}$.  Thus, the
spontaneous symmetry breaking occurs in the D ground state.  This is
consistent with the necessary condition for the appearance of the plateau in
the ground state magnetization curve,
\begin{equation}
  n (S - m) = {\rm integer}\;,
\end{equation}
obtained by Oshikawa, Yamanaka, and Affleck,~\cite{rf:9}  Here, $n$ and $m$
are, respectively, the period of translational symmetry of the plateau state
and the average magnetization per site in the plateau, and $S$ is the
magnitude of spins in the system.  (Note that $n\!=\!6$, $S\!=\!1/2$ and
$m\!=\!0$ for the present ground D state in the $H\!=\!0$ case.)

In the present paper, we discuss the derivation of the ground state phase
diagram in more detail.  Furthermore, we discuss the ground state
magnetization curve, as mentioned before.

\section{Ground State Phase Diagram in the $H\!=\!0$ Case}

Let us denote, respectively, by $E^{(0)}(S_{\rm tot};\,N)$ and
$E^{(1)}(S_{\rm tot};\,N)$ the lowest and second-lowest energy eigenvalues
for a given set of $S_{\rm tot}$ and $N$ of the Hamiltonian
${\cal H}_{\rm ex}$.  We have calculated numerically these eigenvalues for
finite size systems with \hbox{$N\!=\!6$}, $12$, $18$, and $24$, employing the
computer program package KOBEPACK/S~\cite{rf:10} coded by one of the present
authors (M.~K.) by means of the Lancz{\"o}s technique.  Our calculation shows
that, depending on the values of $J_1$, $J_2$, and $J_3$, only
$E^{(0)}(0;\,N)$ or $E^{(0)}(N/6;\,N)$ becomes minimum among
$E^{(0)}(S_{\rm tot}; N)$'s for all values of $S_{\rm tot}(=\!0$, $1$,
$\cdots$, $N/2$).  This is consistent with the fact that the ground state
phase diagram is composed of the D, SF, and FRI phases,~\cite{rf:7} as
discussed in {\S}$\,$1.

The ground state phase is the FRI phase when
$E^{(0)}(N/6;\,\infty)\!<\!E^{(0)}(0;\,\infty)$, and it is the D or the SF
phase when $E^{(0)}(N/6;\,\infty)\!>\!E^{(0)}(0;\,\infty)$.  We estimate the
critical value $J_{i,{\rm c}}^{{\rm FRI-}0}$ of $J_i$ ($i\!=\!1$ $2$, or $3$)
between the former phase and one of the latter phases for a given set of the
other two $J$'s in the following way.  We first calculate numerically the
value $J_{i,{\rm c}}^{{\rm FRI-}0}(N)$ which satisfies
\begin{equation}
  E^{(0)}(N/6;\,N) = E^{(0)}(0;\,N)\;.
\end{equation}
Then, we estimate $J_{i,{\rm c}}^{{\rm FRI-}0}$ by fitting
$J_{i,{\rm c}}^{{\rm FRI-}0}(N)$'s with $12$, $18$, and $24$ to a quadratic
function of $1/N^2$, that is,
\begin{equation}
  J_{i,{\rm c}}^{{\rm FRI-}0}(N)
    = J_{i,{\rm c}}^{{\rm FRI-}0} + \frac{a}{N^2} + \frac{b}{N^4}\;,
\end{equation}
where $a$ and $b$ are numerical constants.

On the other hand, in order to estimate the critical value
$J_{i,{\rm c}}^{\rm D-SF}$ between the D and SF phases, we employ the method
of level spectroscopy~\cite{rf:8,rf:11}, which has been successfully applied
to estimate numerically the BKT critical points in many cases.  The procedure
of the method is as follows.  (For the physical interpretation of this
procedure the reader is referred, for example, to ref.$\,$7.)  First, we
introduce the singlet-singlet energy gap $\Delta_{\rm ss}(N)$ and the
singlet-triplet energy gap $\Delta_{\rm ss}(N)$ for the finite-$N$ system
defined by
\begin{subeqnarray}
  \Delta_{\rm ss} &&\hspace{-0.80truecm}(N) 
               = E^{(1)}(0;\,N) - E^{(0)}(0;\,N)\;,\\
  \Delta_{\rm st} &&\hspace{-0.80truecm}(N) 
               = E^{(0)}(1;\,N) - E^{(0)}(0;\,N)\;,
\end{subeqnarray}
where $E^{(0)}(0;\,N)$ is nothing but the ground state energy.  Then, we
calculate the value $J_{i,{\rm c}}^{\rm D-SF}(N)$ satisfying
\begin{equation}
  \Delta_{\rm ss}(N) = \Delta_{\rm st}(N)\;.
\end{equation}
Finally, we extrapolate the results $J_{i,{\rm c}}^{\rm D-SF}(N)$'s to
$N\!\to\!\infty$ by the use of a polynomial of $1/N^2$ to estimate
$J_{i,{\rm c}}^{\rm D-SF}$.  In the practical calculations  we have performed
this extrapolation by using the values of $J_{i,{\rm c}}^{\rm D-SF}(N)$'s
with $12$, $18$, and $24$ and the formula,
\begin{equation}
  J_{i,{\rm c}}^{{\rm D-SF}}(N)
    = J_{i,{\rm c}}^{\rm D-SF} + \frac{a'}{N^2} + \frac{b'}{N^4}
\end{equation}
with numerical constants $a'$ and $b'$.

We can obtain the ground state phase diagram on a exchange constant parameter
plane by plotting the estimated critical points and by connecting them as
smoothly as possible.  The results are depicted in Figs.$\,$2(a) and 2(b),
which show, respectively, the phase diagram on the $J_3$ versus $J_2$ plane
with $J_1$ fixed at $J_1\!=\!1$ and that on the $J_3$ versus $J_1$ plane with
$J_1\!=\!1$; note that the former is a replot of Fig.$\,$9 in ref.$\,$7.  In
these phase diagrams the case where $J_1\!<\!J_3$ is included to make them
more complete.

\begin{figure}
  \begin{center}
    \psbox[width=7.8cm]{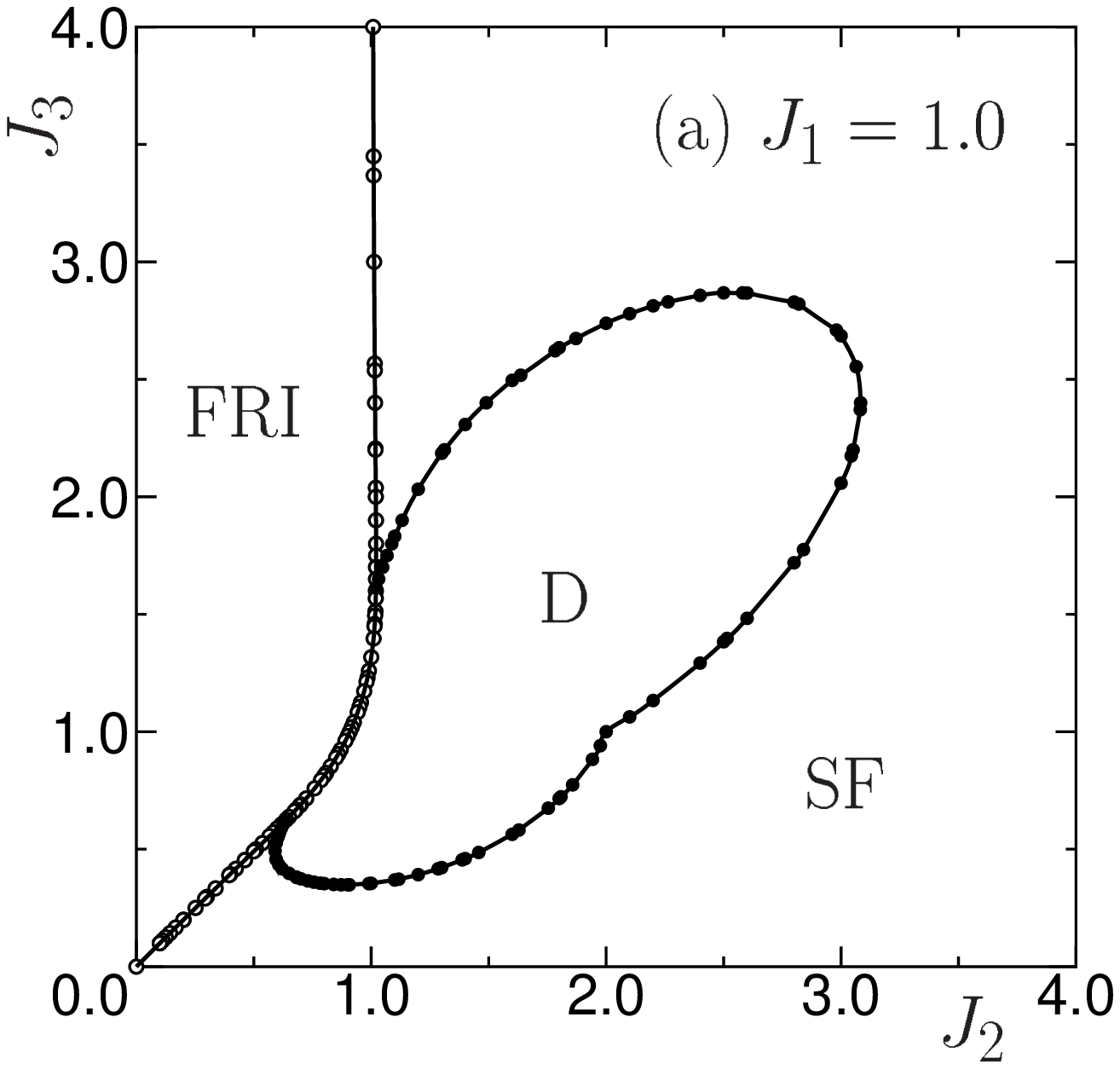}
    \vskip0.8cm
    \psbox[width=7.8cm]{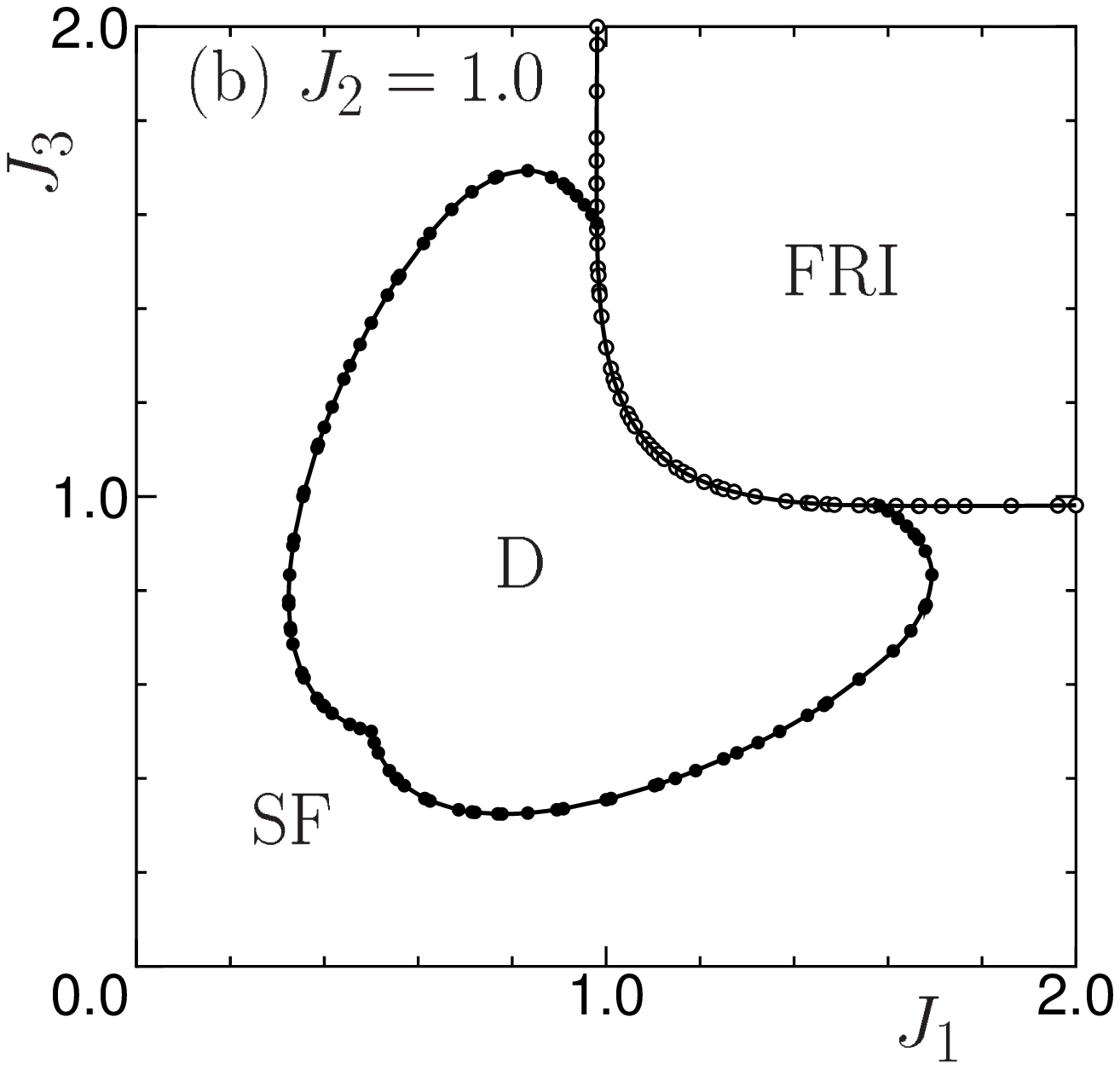}
  \end{center}
\caption{Ground state phase diagrams (a) on the $J_3$ versus $J_2$ plane with
$J_1\!=\!1$ and (b) on the $J_3$ versus $J_1$ plane with $J_1\!=\!1$.  The
phase diagrams are composed of the dimerized (D), spin fluid (SF), and
ferrimagnetic (FRI) phases.}
\end{figure}

The phase transition between the FRI phase and the D or the SF phase is, of
course, of first order.  Furthermore, when $J_1\!=\!J_3$, the transition
between the D and SF phases, or equivalently, that between Takano ${\it et}$
${\it al.}$'s TD and DM phases, at $J_2/J_1\!=\!2$ is also of first
order.  As is shown in Appendix, the ground state energy
$\varepsilon_{\rm g}(N)$ per site for \hbox{$0.909\!<\!J_2/J_1\!<\!2$} and
that for the \hbox{$2\!<\!J_2/J_1$} case are, respectively, given by
$\varepsilon_{\rm g}(N)\!=\!E_{\rm TD}(N)/N\!=\!-(4J_1+J_2)/12$ and
$\varepsilon_{\rm g}(N)\!=\!E_{\rm DM}(N)/N\!=\!-J_2/4$, and therefore,
plotting $\partial\varepsilon_{\rm g}(N)/\partial J_1$ as a function of
$J_1$, we have a discontinuity at $J_1\!=\!J_2/2$.

We assert, for the following reasons, that the first order phase transition is
peculiar only to the above two cases in the present system.  Let us first give
a physical consideration.  It is well known that the ground state of the
$S\!=\!1/2$ chain with uniform, antiferromagnetic nearest and
next-nearest neighbor interactions is a two-fold degenerate dimerized state
which is essentially equivalent to those shown by Figs.$\,$1(c) and 1(d), as
far as the latter interaction is sufficiently larger than the former
one.~\cite{rf:8,rf:12,rf:13}  If a bond alternation is introduced in the
nearest neighbor interaction as a perturbation, the degeneracy of the ground
state is lifted, and one of the dimerized state or the other becomes the
ground state depending upon the value of a bond alternation parameter.  Thus,
sweeping this parameter, we have the first order transition in the ground
state at the uniform point.  In fact, the first order transition has been
observed, for example, in the ground state of a generalized $S\!=\!1/2$ ladder
with additional exchange interactions on diagonal bonds,~\cite{rf:14} which
has bond alternating nearest neighbor interactions when represented in a
one-dimensional chain scheme like Fig.$\,$1(b).  In the present distorted
diamond chain, on the other hand, we have no bond alternation, but have a bond
trimerization in the nearest neighbor interaction [see Fig.$\,$1(d)] which
never lifts a two-fold degeneracy in the D ground state.  Thus, no first order
transition occurs in the ground state of the present chain, except for the
two peculiar cases stated above.


In order to supplement the above reason discussed physically, we have
calculated numerically the $J_1$-dependence of
$\partial\varepsilon_{\rm g}(N)/\partial J_1$ with $N\!=\!6$, $12$, $18$,
and $24$ for the $J_2\!=\!1$ and $J_3/J_1\!=\!0.999$ case, for which the D-SF
critical point is given by $J_{1,{\rm c}}^{\rm D-SF}\!\sim\!0.5002$ according
to the phase diagram shown in Fig.$\,$2(b).  The result is plotted in
Fig.$\,$3, together with that for the $J_2\!=\!1$ and $J_3/J_1\!=\!1$ case
mentioned above.  This figure demonstrates clearly that
$\partial\varepsilon_{\rm g}(N)/\partial J_1$ for the $J_2\!=\!1$ and
$J_3/J_1\!=\!0.999$ case have no anomaly around $J_1\!=\!0.5$.  This result
suggests that we have no first order transition in the $J_1\!<\!J_3$ case
(and also in the $J_1\!>\!J_3$ case) except for the FRI-D transition.

\begin{figure}
  \begin{center}
    \psbox[width=7.8cm]{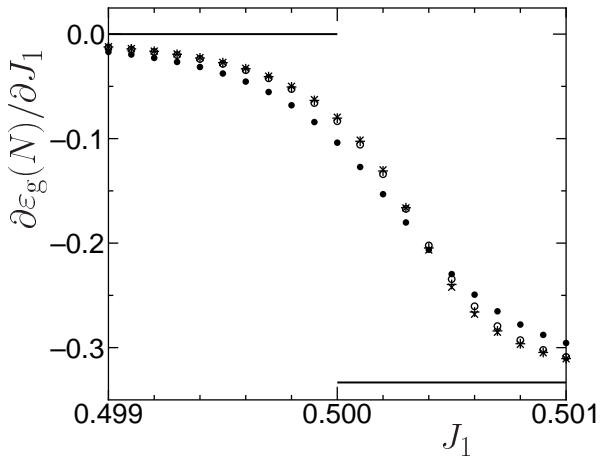}
  \end{center}
\caption{Plot versus $J_1$ of $\partial\varepsilon_{\rm g}(N)/\partial J_1$
with \hbox{$N\!=\!6$} (closed circles), $12$ (open circles), $18$ (pluses),
and $24$ (crosses) for $J_2\!=\!1$ and $J_3/J_1\!=\!0.999$, together with
that (solid lines) for $J_2\!=\!1$ and $J_3/J_1\!=\!1$.}
\end{figure}

\section{Ground State Magnetization Curve}

We have calculated the ground state magnetization curve of the present system,
using the DMRG method proposed originally by White.~\cite{rf:2}  Here,
we describe briefly the procedure for our DMRG calculation.  We employ the
finite system algorithm~\cite{rf:3} as improved by White~\cite{rf:15}, which
reduces substantially the required computational time.  The maximum number of
block states we keep in the calculation is \hbox{$m_{\rm max}\!=\!100$}, which
leads to the truncation error of the order of $10^{-8}$.  Comparing the
calculated results for the cases where \hbox{$m\!=\!70$} and
\hbox{$m\!=\!100$} block states are kept, we may conclude that the results for
\hbox{$m\!=\!100$} are within an accuracy of roughly
$10^{-4}\!\sim\!10^{-5}$.  Thus, we adopt these results as our final results
without carrying out any $m$-extrapolation.  We assume open boundary
conditions for the Hamiltonian ${\cal H_{\rm ex}}$, subtracting the two terms,
$J_1\,\vecS_{N}\cdot\vecS_{N+1}\bigl(\equiv\!J_1\vecS_{N}\cdot\vecS_{1}\bigr)$
and
$J_3\,\vecS_{N}\cdot\vecS_{N+2}\bigl(\equiv\!J_3\vecS_{N}\cdot\vecS_{2}\bigr)$,
from the expression of eq.$\,$(1.1b).  This is because, as is well known, open
boundary conditions make the DMRG method work more effectively.  It should be
noted that these boundary conditions lead to an open chain which has no
inversion symmetry with respect to its center.

For finite size systems with up to $N\!=\!96$ spins, we have calculated the
lowest energy eigenvalue ${\bar E}^{(0)}(M;\,N)$ of ${\cal H_{\rm ex}}$
within the subspace determined by the value $M$ of the $z$-component of
$\vecS_{\rm tot}$.  Once the values of $\bE^{(0)}(M;\,N)$ for
nonnegative $M$'s ($M\!=\!0$, $1$, $\cdots$, $N/2$) are known, the ground
state magnetization curve can be obtained by plotting as a function of $H$
the average magnetization $m(N)(=\!M/N)$ per site determined from
\begin{equation}
 m(N)
   = \cases{ 0 & \hspace{-3.7cm}
               $\bigl($when $\Delta \bE^{(0)}_q(M,M-q;\,N) > H\bigr)\;$,  \cr
            \frac{\max \bigl[\,M\,\big|\,\Delta \bE^{(0)}_q(M,M-q;\,N)
                                                      < H \,\bigr]}{N}
               & (otherwise)$\;$,                                         \cr}
\end{equation}
with $M\!=\!1$, $2$, $\cdots$, $N/2$, and $q\!=\!1$, $2$, $\cdots$, $M$, where
\begin{equation}
 \Delta \bE^{(0)}_q(M,M-q;\,N)
         = \frac{\bE^{(0)}(M;\,N) - \bE^{(0)}(M-q;\,N)}{q}\;.
\end{equation}
It is noted that $m(N)\!=\!0$ when $0\!\leq\!H\!\leq\!H_0(N)$, where
\hbox{$H_0(N)\!=\!\min\bigl\{\Delta \bE^{(0)}_q(q,0;\,N)\bigr\}$}, and the
saturation field $H_{\rm s}(N)$ is given by
\hbox{$H_{\rm s}(N)\!=\!\max\bigl\{\Delta \bE^{(0)}_q(\frac{N}{2},\frac{N}{2}-1;\,N)\bigr\}$}.  The resulting magnetization curve is a stepwisely increasing
function of $H$.  In particular, when the ground state in the case of
\hbox{$H\!=\!0$} is the \hbox{$S_{\rm tot}\!=\!0$} state and the competition
among the three interactions is not too strong, the magnetization curve has
$N/2$ steps, starting from \hbox{$m(N)\!=\!0$} and ending at
\hbox{$m(N)\!=\!1/2$}; at each step $m(N)$ increases by
$1/N$.~\cite{rf:16}  Then,
$H_0(N)\!=\!\Delta \bE^{(0)}_1(1,0;\,N)\!=\!\Delta_{\rm st}(N)$, and
$H_{\rm s}(N)\!=\!\Delta \bE^{(0)}_1(M,\hbox{$M\!-\!1;$}\,N)$.  Following
Bonner and Fisher's pioneering work~\cite{rf:17}, we may obtain, except for
plateau regions, a satisfactorily good approximation to the magnetization
curve in the thermodynamic (\hbox{$N\!\to\!\infty$}) limit by drawing a
smooth curve through the midpoints of the steps in the finite size results.

\begin{figure}
  \begin{center}
    \psbox[width=7.8cm]{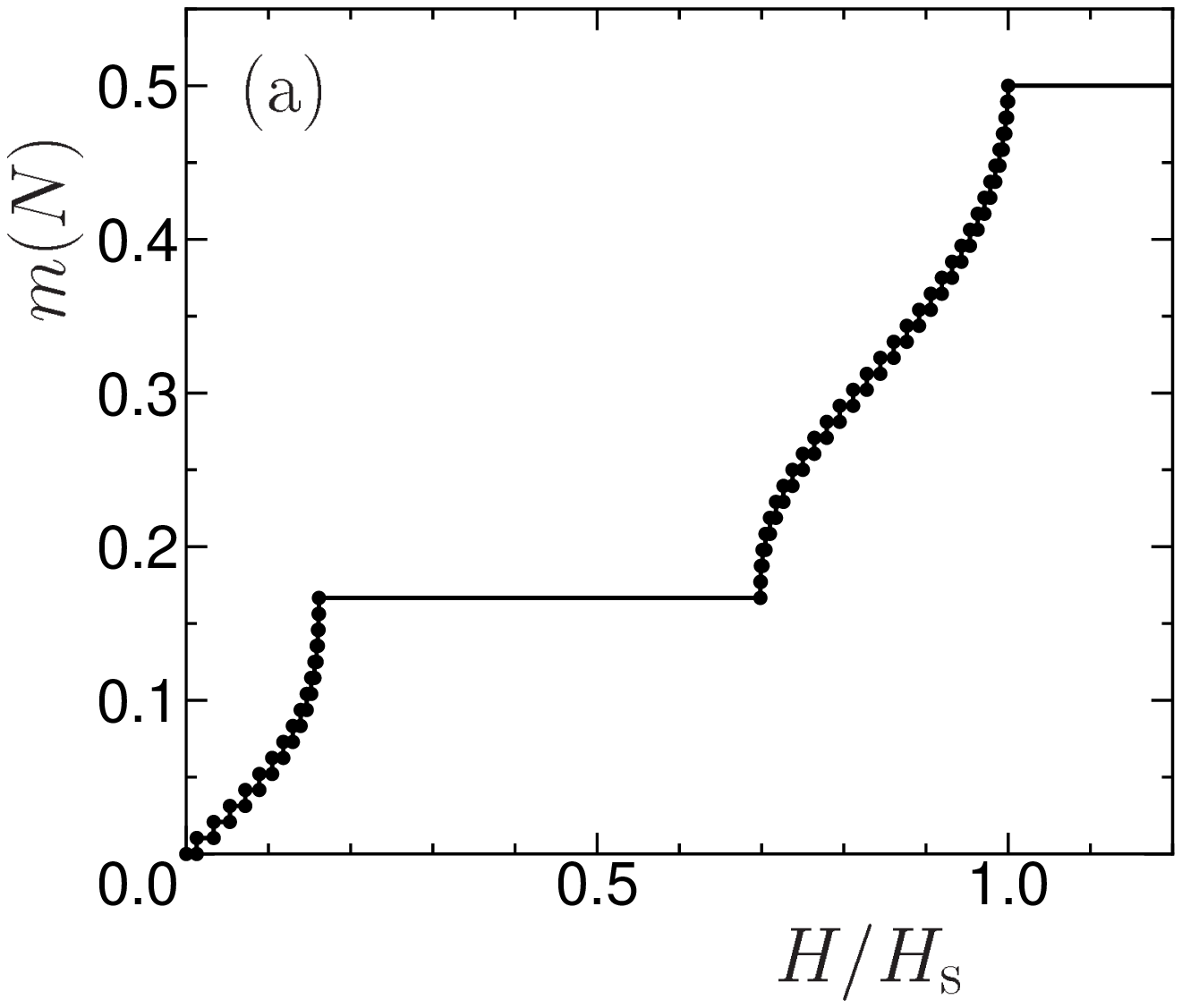}
    \vskip0.8cm
    \psbox[width=7.8cm]{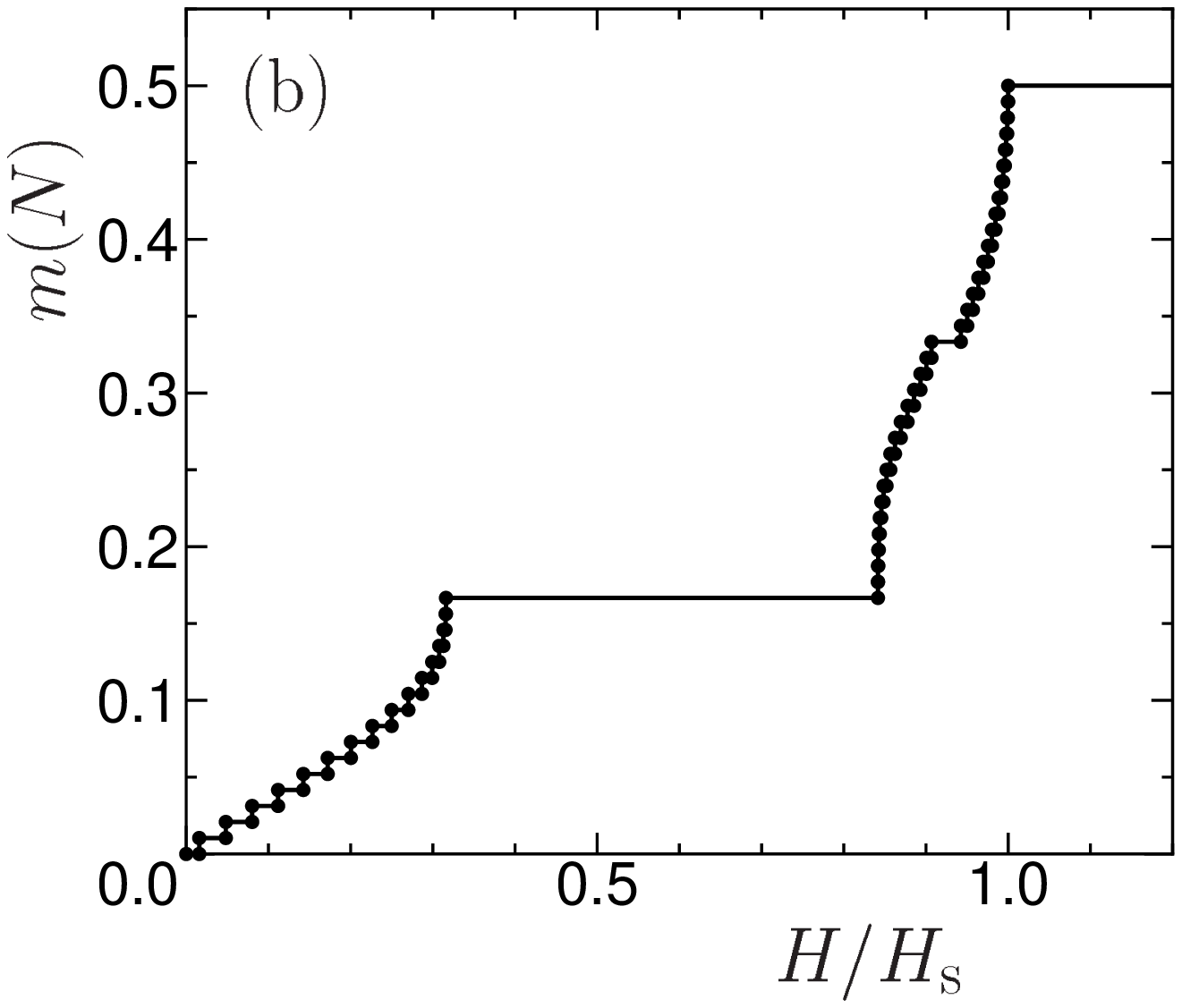}
  \end{center}
\caption{Ground state magnetization curves calculated for $N\!=\!96$
(a) in the case~(a) where $J_1\!=\!1.0$, $J_2\!=\!0.8$, and $J_3\!=\!0.5$
and (b) in the case~(b) where $J_1\!=\!1.0$, $J_2\!=\!0.8$, and
$J_3\!=\!0.3$.}
\end{figure}

\begin{table}
\caption{Numerical values of $H_0(N)/H_{\rm s}(N)$,
$H_{1/3,{\rm l}}(N)/H_{\rm s}(N)$, $H_{1/3,{\rm h}}(N)/H_{\rm s}(N)$,
$H_{2/3,{\rm l}}(N)/H_{\rm s}(N)$, and $H_{2/3,{\rm h}}(N)/H_{\rm s}(N)$
estimated in the cases~(a) and (b).  Note that the figures in the parentheses
for the \hbox{$N\!\to\!\infty$} values show errors in the last digit.}
\label{table:1}
\vspace{0.3truecm}
(a)~the case (a) where $J_1\!=\!1.0$, $J_2\!=\!0.8$, and $J_3\!=\!0.5$.
\vspace{-0.2truecm}
\begin{tabular}{@{\hspace{\tabcolsep}\extracolsep{\fill}}ccccc}
\hline
     & $N\!=\!48$  &  $N\!=\!72$  &  $N\!=\!96$  &  $N\!\to\!\infty$   \\
\hline
     $\frac{H_0(N)}{H_{\rm s}(N)}$
     &  $0.024525$  &  $0.016736$  &  $0.012795$  &  $0.0033\,(5)$     \\
     $\frac{H_{1/3,{\rm l}}(N)}{H_{\rm s}(N)}$
     &  $0.160260$  &  $0.160875$  &  $0.161043$  &  $0.1612\,(1)$     \\
     $\frac{H_{1/3,{\rm h}}(N)}{H_{\rm s}(N)}$
     &  $0.702154$  &  $0.699845$  &  $0.699090$  &  $0.6982\,(1)$     \\
\hline
\end{tabular}
\vspace{0.3truecm}
(b)~the case (b) where $J_1\!=\!1.0$, $J_2\!=\!0.8$, and $J_3\!=\!0.3$.
\vspace{-0.2truecm}
\begin{tabular}{@{\hspace{\tabcolsep}\extracolsep{\fill}}ccccc}
\hline
     & $N\!=\!48$  &  $N\!=\!72$  &  $N\!=\!96$  &  $N\!\to\!\infty$   \\
\hline
     $\frac{H_0(N)}{H_{\rm s}(N)}$
     &  $0.031482$  &  $0.021075$  &  $0.015841$  &  $0.0        $     \\
     $\frac{H_{1/3,{\rm l}}(N)}{H_{\rm s}(N)}$
     &  $0.312021$  &  $0.314803$  &  $0.315499$  &  $0.3162\,(1)$     \\
     $\frac{H_{1/3,{\rm h}}(N)}{H_{\rm s}(N)}$
     &  $0.842373$  &  $0.841792$  &  $0.841618$  &  $0.8414\,(1)$     \\
     $\frac{H_{2/3,{\rm l}}(N)}{H_{\rm s}(N)}$
     &  $0.902840$  &  $0.905461$  &  $0.906667$  &  $0.9084\,(1)$     \\
     $\frac{H_{2/3,{\rm h}}(N)}{H_{\rm s}(N)}$
     &  $0.944329$  &  $0.942955$  &  $0.942391$  &  $0.9416\,(1)$     \\
\hline
\end{tabular}
\end{table}

Figure 4 shows the magnetization curves calculated for $N\!=\!96$;
Figs.$\,$4(a) and 4(b) are, respectively, for the case~(a) where
$J_1\!=\!1.0$, $J_2\!=\!0.8$, and $J_3\!=\!0.5$, and for the case~(b) where
$J_1\!=\!1.0$, $J_2\!=\!0.8$, and $J_3\!=\!0.3$.  In each case, we have
a $1/3$-plateau, and moreover, in the case~(b), we have a $2/3$-plateau in
addition to this.  We denote the highest and lowest values of $H$ giving the
$p$-plateau ($p\!=\!1/3$, $2/3$) by $H_{{p},{\rm h}}(N)$ and
$H_{{p},{\rm l}}(N)$, respectively.  We have calculated these as well as
$H_0(N)$ and $H_{\rm s}(N)$ also for \hbox{$N\!=\!72$} and $48$ in the
cases~(a) and (b), and have extrapolated the results to
\hbox{$N\!\to\!\infty$} to estimate the values in the thermodynamic
limit.  The results are tabulated in Table~\ref{table:1}.  We see from this
Table that, except for $H_0(N)$, the finite size results for
\hbox{$N\!=\!96$} give good approximations to the results in the thermodynamic
limit.

\begin{figure}
  \begin{center}
    \psbox[width=7.8cm]{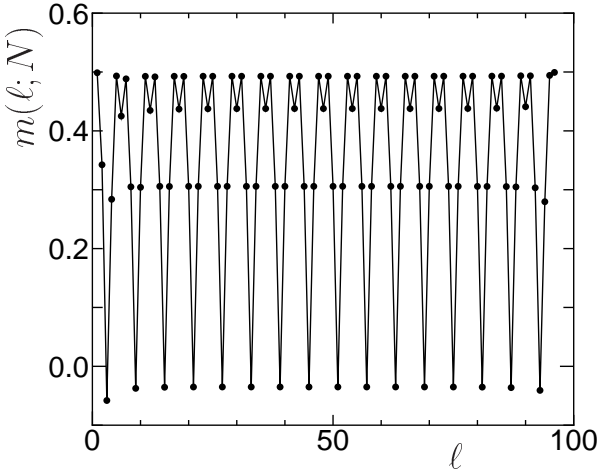}
  \end{center}
\caption{Plot versus $\ell$ of $m(\ell;\,N)$ with $N\!=\!96$ for the
2/3-plateau state in the case~(b) where $J_1\!=\!1.0$, $J_2\!=\!0.8$,
and $J_3\!=\!0.3$.  Solid lines are guides to the eye.}
\end{figure}

The 1/3-plateau state is essentially the same as the ferrimagnetic state
which becomes the ground state when
\hbox{$J_3\!>\!J_{3,{\rm c}}^{{\rm FRI-}0}\!\sim\!0.813$} in the
\hbox{$J_1\!=\!1.0$}, \hbox{$J_2\!=\!0.8$}, and \hbox{$H\!=\!0$} case.  On
the other hand, we see from eq.$\,$(1.2) that the periodicity $n$ of the
2/3-plateau state should be a multiple of six, since \hbox{$S\!=\!1/2$} and
\hbox{$M\!=\!1/3$}.  In order to check this fact, we have calculated, by
means of the DMRG method, the $\ell$-dependence of the expectation value
$m(\ell;\,N)$ of $S_\ell^z$ for the 2/3-plateau state in the case~(b).  The
results for \hbox{$N\!=\!96$} depicted in Fig.$\,$5 clearly demonstrates that
\hbox{$n\!=\!6$} for this state, which means that the translational
symmetry is spontaneously broken also in this state as in the D state.  Thus,
the mechanism for the appearance of the 2/3-plateau in the present case is
considered to be the same as that, which has been clarified by
Totsuka,~\cite{rf:18} for the appearance of the 1/2-plateau in the ground
state magnetization curve of the $S\!=\!1/2$ antiferromagetic chain with bond
alternating nearest and uniform next-nearest neighbor
interactions.~\cite{rf:19}  Based on this mechanism, we can determine, by
using the method of level spectroscopy~\cite{rf:8,rf:11}, the region, where
the 2/3-plateau appears, of $J_i$ for a given set of the other two
$J$'.  Leaving the details of this investigation for a forthcoming paper, we
only mention here that, according our preliminary result, the 2/3-plateau
appears when $0.169\lsim J_3\lsim0.375$ in the $J_1\!=\!1.0$ and
$J_2\!=\!0.8$ case.

\section{Concluding Remarks and Discussion}

We have explored the ground state properties of an \hbox{$S\!=\!1/2$}
distorted diamond chain described by the Hamiltonian ${\cal H}$ [see
eqs.$\,$(1.1a)-(1.1c)], which models well a trimerized \hbox{$S\!=\!1/2$}
spin chain system Cu$_3$Cl$_6$(H$_2$O)$_2$$\cdot$2H$_8$C$_4$SO$_2$.  Using an
exact diagonalization method by means of the Lancz{\"o}s technique, we have
determined the ground state phase diagram in the \hbox{$H\!=\!0$}
case [Figs.$\,$2(a) and 2(b)], composed of the D, SF, and FRI phases.  As has
been mentioned in {\S}$\,$1, the ground state of
Cu$_3$Cl$_6$(H$_2$O)$_2$$\cdot$2H$_8$C$_4$SO$_2$ is nonmagnetic with a
relatively small but finite energy gap.  Furthermore, the three exchange
constants, $J_1$, $J_2$, and $J_3$, in this system satisfy
$J_1\!>\!J_2,~J_3\!>\!0$.  From these facts together with the obtained phase
diagram shown in Figs.$\,$2(a), we may anticipate that in
Cu$_3$Cl$_6$(H$_2$O)$_2$$\cdot$2H$_8$C$_4$SO$_2$ the values of $J_2/J_1$ and
$J_3/J_1$ are in the region of $0.7\lsim J_2/J_1\lsim0.9$ and
$0.45\lsim J_3/J_1\lsim0.65$.  We have also calculated, by the use of the
density matrix renormalization group method, the ground state magnetization
curve for the case~(a) where $J_1\!=\!1.0$, $J_2\!=\!0.8$, and $J_3\!=\!0.5$
[Fig.4$\,$(a)], and the case~(b) where $J_1\!=\!1.0$, $J_2\!=\!0.8$, and
$J_3\!=\!0.3$ [Fig.$\,$4(b)].  We have found that in the case~(b) the
$2/3$-plateau appears in addition to the $1/3$-plateau which also appears in
the case~(a), and have clarified that the translational symmetry of the
Hamiltonian ${\cal H}_{\rm ex}$ is spontaneously broken in the $2/3$-plateau
state.

Figure 2(a) demonstrates that the boundary line between FRI and SF phases in
the limit of \hbox{$0\!\leq\!J_2\!\ll\!J_1$} and
\hbox{$0\!\leq\!J_3\!\ll\!J_1$} is approximately given by
\hbox{$J_2\!=\!J_3$}.  This can be understood analytically by performing a
degenerate perturbation calculation around the truncation point
\hbox{$J_2\!=\!J_3\!=\!0$} in the following way.  The unperturbed Hamiltonian
${\cal H}_0$ is given by the sum of the Hamiltonian $h_\ell$ for the $\ell$th
trimer consisting of $\vecS_{3\ell-1}$, $\vecS_{3\ell}$, and
$\vecS_{3\ell+1}$:
\begin{subeqnarray}
 {\cal H}_0 =&&\hspace{-0.80truecm}\sum_{\ell=1}^{N/3} h_\ell\,,  \\
    &&\hspace{-0.80truecm}
      h_\ell = \vecS_{3\ell-1} \cdot \vecS_{3\ell}
                + \vecS_{3\ell} \cdot \vecS_{3\ell+1}\,,
\end{subeqnarray}
where the value of $J_1$ is taken to be \hbox{$J_1\!=\!1.0$} as the unit of
energy.  The eigenfunctions, $\phi_\ell^{(1)}$ and $\phi_\ell^{2}$, of the
lowest energy states of $h_\ell$, which are two-fold degenerate, are
expressed, by the use of $\a_\ell$ and $\b_\ell$ representing, respectively,
the $S_\ell^z\!=\!1/2$ and $S_\ell^z\!=\!-1/2$ single spin states, as
\begin{subeqnarray}
 &&\hspace{-0.8truecm} \phi_\ell^{(1)}
   = \frac{1}{\sqrt{6}}
       \bigl\{ \vert \a_{3\ell-1} \a_{3\ell} \b_{3\ell+1} \rangle
         - 2\, \vert \a_{3\ell-1} \b_{3\ell} \a_{3\ell+1} \rangle 
                                                                \nonumber  \\
 &&\hspace{2.5truecm}
         +  \, \vert \b_{3\ell-1} \a_{3\ell} \a_{3\ell+1} \rangle \bigl\}\;,
                                                                           \\
 &&\hspace{-0.8truecm} \phi_\ell^{(2)}
   = \frac{1}{\sqrt{6}}
       \bigl\{ \vert \b_{3\ell-1} \b_{3\ell} \a_{3\ell+1} \rangle
         - 2\, \vert \b_{3\ell-1} \a_{3\ell} \b_{3\ell+1} \rangle
                                                                \nonumber  \\
 &&\hspace{2.5truecm}
         +  \, \vert \a_{3\ell-1} \b_{3\ell} \b_{3\ell+1} \rangle \bigl\}\;.
\end{subeqnarray}
Restricting only to these two eigenfunctions, we can represent
$\vecS_{3\ell-1}$, $\vecS_{3\ell}$, and $\vecS_{3\ell+1}$ in terms of the
pseudo \hbox{$S\!=\!1/2$} operator $\vecT_\ell$ associated with the $\ell$th
trimer:
\begin{subeqnarray}
 &&\hspace{-0.8truecm}
           S_{3\ell-1}^\pm = - \frac{2}{3} T_\ell^\pm\;,
    \qquad S_{3\ell-1}^z   =   \frac{2}{3} T_\ell^z\;,     \\
 &&\hspace{-0.8truecm}
           S_{3\ell  }^\pm =   \frac{1}{3} T_\ell^\pm\;,
    \qquad\hspace{0.6truecm}
           S_{3\ell  }^z   = - \frac{1}{3} T_\ell^z\;,     \\
 &&\hspace{-0.8truecm}
           S_{3\ell+1}^\pm = - \frac{2}{3} T_\ell^\pm\,,
    \qquad S_{3\ell+1}^z   =   \frac{2}{3} T_\ell^z\;,
\end{subeqnarray}
where the \hbox{$T_\ell^z\!=\!1/2$} and \hbox{$T_\ell^z\!=\!-1/2$} states
correspond to $\phi_\ell^{(1)}$ and $\phi_\ell^{2}$,
respectively.  Substituting eqs.$\,$(4.3a)-(4.3c) into the second ($J_2$) and
third ($J_3$) terms in the right-hand side of eq.$\,$(1.1b) leads to the
following effective Hamiltonian ${\cal H}_{\rm eff}$:
\begin{equation}
  {\cal H}_{\rm eff}
    = \frac{4}{9}\;(J_2 - J_3)
             \sum_{\ell=1}^{N/3} \vecT_\ell \cdot \vecT_{\ell+1}
\end{equation}
with \hbox{$\vecT_{(N/3)+1}\!\equiv\!\vecT_1$}.  We see from this equation
that the ground state of the pseudo $\vecT_\ell$ spin system is the
ferromagnetic or the SF state depending upon whether $J_2\!>\!J_3$ or
$J_2\!<\!J_3$, the former and the latter corresponding, respectively, to the
FRI and SF states in the original $\vecS_\ell$ spin system.  Thus, it is shown
that the $J_2\!=\!J_3$ line yields the boundary line between FRI and SF phases
in the limit of $0\!\leq\!J_2\!\ll\!J_1$ and $0\!\leq\!J_3\!\ll\!J_1$.

\section*{Acknowledgments}
We would like to thank Professors H.-J.~Mikeska, K.~Takano, K.~Kubo, and
H.~Tanaka for valuable discussions.  We are also thankful to the Supercomputer
Center, Institute for Solid State Physics, University of Tokyo, the Computer
Center, Tohoku University, and the Computer Room, Yukawa Institute for
Theoretical Physics, Kyoto University for computational facilities.  The
present work has been supported in part by a Grant-in-Aid for Scientific
Research (C) from the Ministry of Education, Science, Sports and Culture.

\appendix
\section{}
Here, we discuss the \hbox{$J_1\!=\!J_3$} case~\cite{rf:6}.  Let us introduce
the following states:
\begin{subeqnarray}
 [\,\ell,\,&&\hspace{-0.8truecm}\ell+1,\,\ell+2,\,\ell+3\,]     \nonumber  \\
   &&\hspace{-0.8truecm}
    = \frac{1}{\sqrt{12}} \bigl\{
             \vert \a_{\ell} \a_{\ell+1} \b_{\ell+2} \b_{\ell+3} \rangle
        +    \vert \b_{\ell} \b_{\ell+1} \a_{\ell+2} \a_{\ell+3} \rangle
                                                                \nonumber  \\
   &&\hspace{0.17truecm}
        + \, \vert \a_{\ell} \b_{\ell+1} \a_{\ell+2} \b_{\ell+3} \rangle
        +    \vert \b_{\ell} \a_{\ell+1} \b_{\ell+2} \a_{\ell+3} \rangle
                                                                \nonumber  \\
   &&\hspace{0.17truecm}
       - \,2 \,\vert \a_{\ell} \b_{\ell+1} \b_{\ell+2} \a_{\ell+3} \rangle
       -   2 \,\vert \b_{\ell} \a_{\ell+1} \a_{\ell+2} \b_{\ell+3} \rangle
                           \bigl\} \;,                          \nonumber  \\
                                                                           \\
 \hspace{-0.5truecm}[\,\ell,\,&&\hspace{-0.8truecm}\ell+1\,]
    = \frac{1}{\sqrt{2}} \bigl\{
             \vert \a_{\ell} \b_{\ell+1} \rangle
      -      \vert \b_{\ell} \a_{\ell+1} \rangle
                           \bigl\} \;,                                     \\
 \hspace{-0.5truecm}[\,\ell\,]
        &&\hspace{-0.8truecm} = \vert \a_{\ell} \rangle
       \qquad {\rm or} \qquad   \vert \b_{\ell} \rangle\,.
\end{subeqnarray}
Then, the eigenfunctions $\Phi_{\rm TD}^\pm(N)$ of the TD state for the
finite-$N$ system is given by
\begin{equation}
  \Phi_{\rm TD}^\pm(N) 
      = \frac{1}{\sqrt{2}}\big\{\Phi_{\rm TD}^{(1)}(N)
                            \pm \Phi_{\rm TD}^{(2)}(N)\big\}
\end{equation}
with
\begin{subeqnarray}
  \Phi_{\rm TD}^{(1)}(N)&&\hspace{-0.8truecm}                     \nonumber \\
    = [&&\hspace{-0.8truecm}1,\,2\,]\,[\,3,\,4,\,5,\,6\,]\,\cdots \nonumber \\
           &&\hspace{-0.8truecm}
                 [\,N\!-5,\,N\!-4\,]\,[\,N\!-3,\,N\!-2,\,N\!-1,\,N\,]\;,
                                                                  \nonumber \\
                                                                            \\
  \Phi_{\rm TD}^{(2)}(N)&&\hspace{-0.8truecm}                     \nonumber \\
    = [&&\hspace{-0.8truecm}N,\,1,\,2,\,3\,]\,[\,4,\,5\,]
                                                         \,\cdots \nonumber \\
           &&\hspace{-0.8truecm}
                 [\,N\!-6,\,N\!-5\,]\,[\,N\!-4,\,N\!-3,\,N\!-2,\,N\!-1\,]\;.
                                                                  \nonumber \\
\end{subeqnarray}
Note here that
$\langle\Phi_{\rm TD}^{(1)}(N)\vert\Phi_{\rm TD}^{(2)}(N)\rangle\!=\!0$.  On
the other hand, the eigenfunction $\Phi_{\rm DM}(N)$ of the DM state for the
finite-$N$ system is
\begin{equation}
   \Phi_{\rm DM}(N)
          = [\,1,\,2\,]\,[\,3\,]\,\cdots\,[\,N\!-3,\,N\!-2\,]\,[\,N\,]\;.
\end{equation}
It is easy to show that the energies of the TD and DM states for the
finite-$N$ system are given, respectively, by
$E_{\rm TD}(N)\!=\!-\frac{N}{12}(4J_1\!+\!J_2)$ and
$E_{\rm DM}(N)\!=\!-\frac{N}{4}J_2$.  We note that $E_{\rm DM}(N)$ is the
total energy of $\frac{N}{3}$ independent singlet dimers mentioned in
{\S}$\,$1, while $E_{\rm TD}(N)$ is the total energy of $\frac{N}{6}$
independent singlet tetramers plus that of $\frac{N}{6}$ independent singlet
dimers.  Furthermore, it is also straightforward to calculate the spin pair
correlation functions in the TD and DM states, defined, respectively, by
\begin{subeqnarray}
  \hspace{-0.7truecm}\omega_{\rm TD}(\ell,\ell';\,N)
     &&\hspace{-0.8truecm}=
       \langle\Phi_{\rm TD}^\pm(N)\vert
          \vecS_\ell\cdot\vecS_{\ell'}\vert\Phi_{\rm TD}^\pm(N)\rangle\;,  \\
  \hspace{-0.7truecm}\omega_{\rm DM}(\ell,\ell';\,N)
     &&\hspace{-0.8truecm}=
       \langle\Phi_{\rm DM}^\pm(N)\vert
          \vecS_\ell\cdot\vecS_{\ell'}\vert\Phi_{\rm DM}^\pm(N)\rangle\;.
\end{subeqnarray}
When \hbox{$N\!=\!12$}, $18$, $\cdots$, the function
$\omega_{\rm TD}(\ell,\ell';\,N)$ for $\ell\!\leq\!\ell'\,$~\cite{rf:20} takes
nonzero values independently of $N$ only in the following cases:
\begin{subeqnarray}
   \omega_{\rm TD}(\ell,\ell;\,N)
      &&\hspace{-0.8truecm}= \frac{3}{4}\;,                                \\
   \omega_{\rm TD}(3\ell-2,3\ell;\,N)
      &&\hspace{-0.8truecm}= \omega_{\rm TD}(3\ell,3\ell+2;\,N)  \nonumber \\
      &&\hspace{-0.8truecm}= - \frac{1}{12}\;,                             \\
   \omega_{\rm TD}(3\ell-1,3\ell;\,N)
      &&\hspace{-0.8truecm}= \omega_{\rm TD}(3\ell,3\ell+1;\,N)  \nonumber \\
      &&\hspace{-0.8truecm}= - \frac{1}{12}\;,                             \\
   \omega_{\rm TD}(3\ell+1,3\ell+2;\,N)
      &&\hspace{-0.8truecm}= - \frac{1}{12}\;,                             \\
   \omega_{\rm TD}(3\ell,3\ell+3;\,N)
      &&\hspace{-0.8truecm}= \frac{1}{24}\;.
\end{subeqnarray}
On the other hand, the nonzero values of $\omega_{\rm DM}(\ell,\ell';\,N)$
for $\ell\!\leq\!\ell'\,$~\cite{rf:20} are
\begin{subeqnarray}
   \omega_{\rm DM}(\ell,\ell;\,N)
      &&\hspace{-0.8truecm}= \frac{3}{4}\;,                                \\
   \omega_{\rm DM}(3\ell+1,3\ell+2;\,N)
      &&\hspace{-0.8truecm}= - \frac{3}{4}\;,
\end{subeqnarray}
which are independent of $N(=\!6$, $12$, $18$, $\cdots)$.






\begin{thebibliography}{99}
\bibitem{rf:16} When the ground state in the case of $H\!=\!0$ is the
$S_{\rm tot}\!=\!N/6$ (ferrimagnetic) state, the magnetization curve starts,
of course, from $m(N)\!=\!N/6$, and when the competition among the three
interactions is sufficiently strong, some of the $N/2$ steps may be skipped.
\end{thebibliography}
\end{document}